\DeclareMathAlphabet{\mathcal}{OMS}{cmsy}{m}{n}
\SetMathAlphabet{\mathcal}{bold}{OMS}{cmsy}{b}{n}

\documentclass[fleqn,10pt]{wlscirep}
\usepackage[utf8]{inputenc}
\usepackage[T1]{fontenc}
\title{Seismic Traveltime Inversion with Quantum Annealing}

\author[1,*]{Hoang Anh Nguyen}
\author[1]{Ali Tura}
\affil[1]{Department of Geophysics, Colorado School of Mines, Golden, 80401, Colorado, USA}

\affil[*]{hoanganh\_nguyen@mines.edu}


\begin{abstract}
This study demonstrates the application of quantum computing based quantum annealing to seismic traveltime inversion, a critical approach for inverting highly accurate velocity models. The seismic inversion problem is first converted into a Quadratic Unconstrained Binary Optimization problem, which the quantum annealer is specifically designed to solve. We then solve the problem via quantum annealing method. The inversion is applied on a synthetic velocity model, presenting a carbon storage scenario at depths of 1000-1300 meters. As an application example, we also show the capacity of quantum computing to handle complex, noisy data environments. This work highlights the emerging potential of quantum computing in geophysical applications, providing a foundation for future developments in high-precision seismic imaging.
\end{abstract}
\begin{document}

\flushbottom
\maketitle

\thispagestyle{empty}

\section*{Introduction}

Quantum computing is an emerging field with significant promise for various scientific and engineering disciplines. As we stand at the frontier of this technological revolution, early-stage research in quantum computing is crucial for the advancement of geophysics. Numerous studies have begun to explore the integration of quantum computing within this field, highlighting its immense and revolutionary potential \cite{moradi18}. For instance, quantum annealers can perform well in solving tomography optimization problems \cite{sarkar18}.  The quantum computing is applied for binary-value full waveform inversion, addressing issues related to velocity variations \cite{greer20}. In the frequency domain, the seismic wave equation can be reduced to a system of linear equations, allowing for the application of quantum annealing \cite{kumar22}. Furthermore, it has been shown that quantum annealing impedance inversion with L1 norm regularization can dramatically enhance accuracy and anti-noise capabilities \cite{wang24}.

A quantum annealer is a specific type of quantum computer designed to solve optimization problems \cite{Yulianti22}. 
The quantum annealing process in quantum annealers can find the minimum energy state of a system, corresponding to the optimal solution of a given problem \cite{MCGEOCH20}. This process is achieved by utilizing quantum fluctuations, allowing the system to tunnel through energy barriers \cite{crosson16}. While there are various types of models in quantum computing \cite{nimbe21,Lu2023}, this particular feature allows quantum annealing to efficiently explore complex energy landscapes, making them particularly well-suited for solving optimization problems. 

Most previous attempts to address seismic problems using quantum annealers have primarily involved relatively simple models \cite{Alulaiw15, Albino22}. For conventional approach by classical computers, the cross-well seismic inversion between boreholes can be computationally expensive \cite{mcmechan83}, necessitating the development of new methods to tackle these challenges. Therefore, in this study, we aim to advance this line of research by applying quantum annealing to a complex problem: Seismic traveltime inversion of the velocity model between two boreholes. Our focus is on developing an inversion strategy that can accurately invert the velocity model with noisy data despite the limitation of the quantum hardware, specifically targeting carbon storage scenarios at depths of 1000-1300 meters. We use quantum annealer at D-Wave Advantage System, which has at least 5000 qubits \cite{mcgeoch2020d}. Clearly, this travel-time inversion method can be applied to other acquisition geometries and data such as surface seismic, vertical seismic profile (VSP), earthquake or micro seismic data.
\section*{Results}

\begin{figure}[!ht]
\centering
\includegraphics[width=0.9\linewidth]{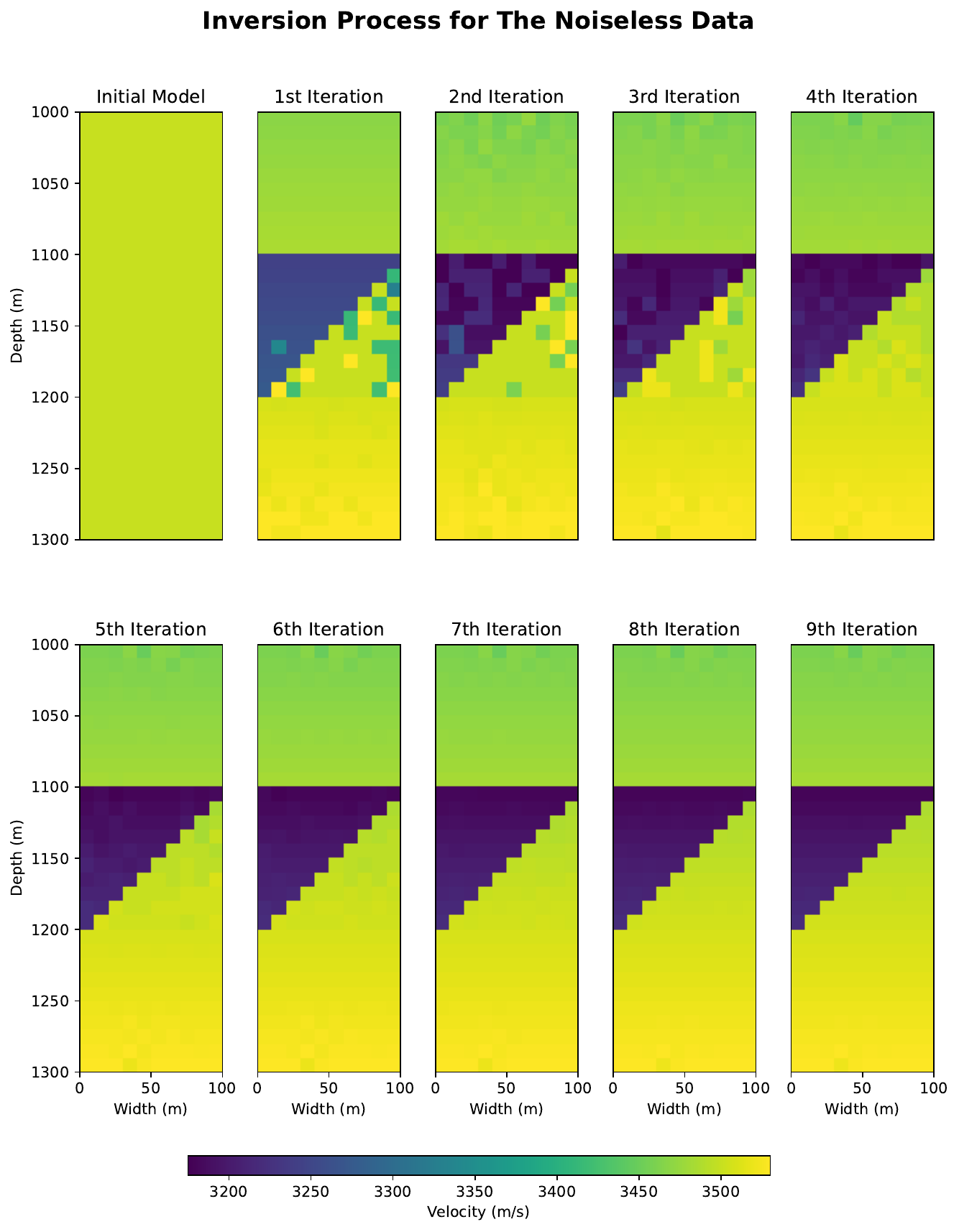}
\caption{The starting model \(v_{ini}\) and the inverted velocity model \(v_{inv}\) over the first 9 iterations with exact, noise-free traveltime data.}
\label{fig:result10loops}
\end{figure}

We start the quantum annealing inversion process with exact traveltime data without noise and constant initial velocity model \(v_{ini}\) of 3475 m/s. The initial model and the results of the inverted model \(v_{inv}\) at each iteration obtained after the first 9 iterations indicate rapid convergence (Fig. \ref{fig:result10loops}). Notably, in the first iteration, the carbon storage area is immediately identified with high precision.

\begin{figure}[!ht]
\centering
\includegraphics[width=0.9\linewidth]{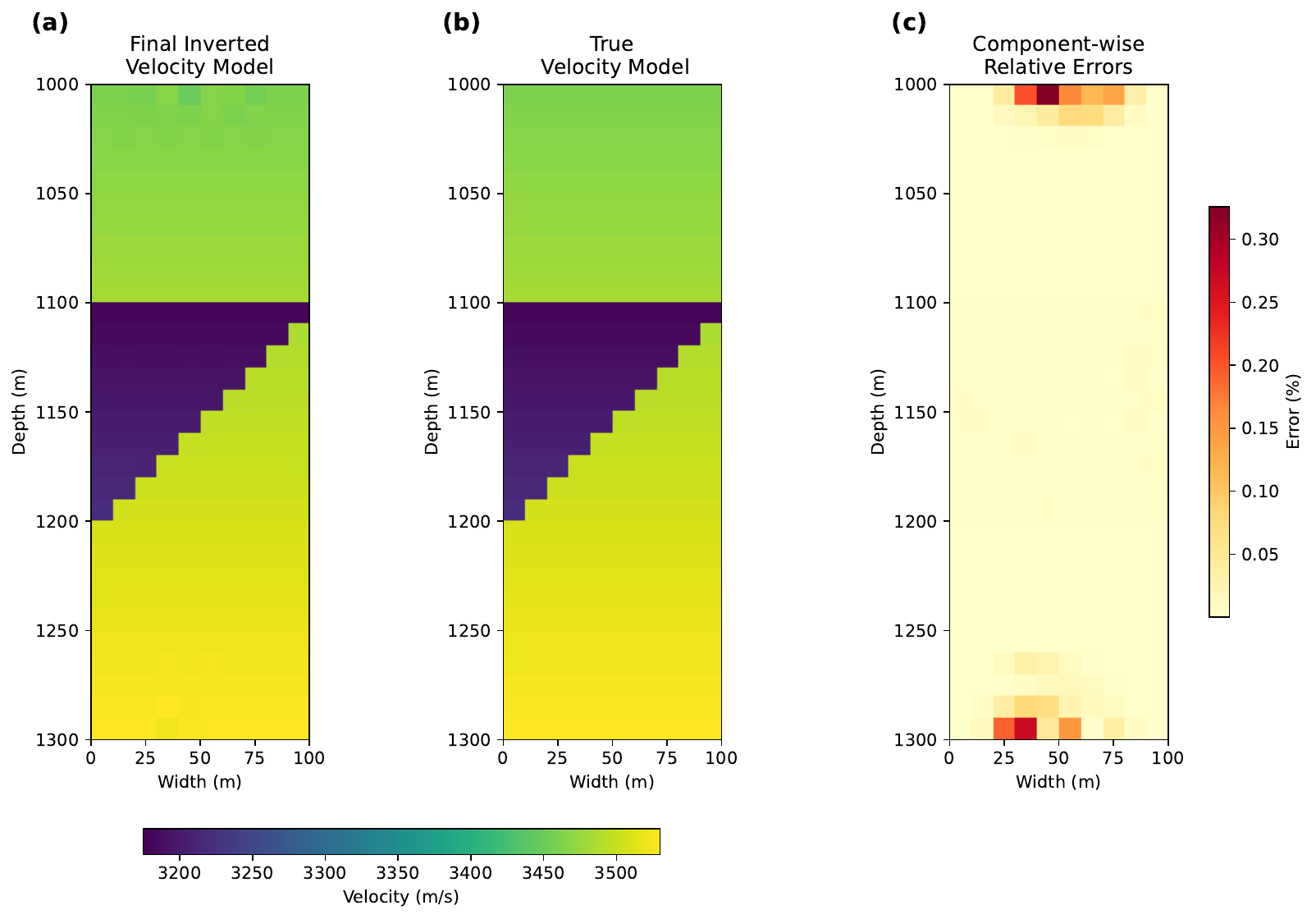}
\caption{Velocity models and error: (a) final noiseless inverted model after 10 iterations \(v_{inv, ij}\), (b) true model \(v_{true, ij}\), and (c) the component-wise relative errors \(e_{ij}\) between the final inverted and true velocity model.}
\label{fig:result_diff}
\end{figure}

The component-wise relative errors \({e}_{ij}\) between the true \(v_{true, ij}\) and final inverted velocity model \(v_{final, ij}\) after 10 iterations is shown in Fig. \ref{fig:result_diff}. The component-wise relative errors are calculated by \({e}_{ij} = |v_{inv, ij} - v_{true, ij}| / |v_{true, ij}|\). The most significant errors occurs in the shallow and deep regions with weakest constraints, yielding a maximum relative error value of about 0.326\%. In contrast, the carbon storage area, spanning depths from 1100 to 1200 m, demonstrates exceptionally low errors due to the high ray coverage. The high-accuracy result underscores the effectiveness of the quantum annealing approach to the traveltime inversion.

For seismic traveltime inversion problem, the quantum annealing method and the classical linear least squares approach produce similar levels of error under ideal conditions, where data is free of noise \cite{Souza22}. However, real-world data often contain random noise, making it essential to assess the robustness of these methods under realistic conditions. Since we use the first-arrival traveltime, the data is relatively clean \cite{Daley2008}. Therefore, we introduce the random noise in a range from 1\% to 5\% into the synthetic data. We compare the outcomes of the Tikhonov regularization least squares, serving as the standard method, with that of the quantum annealing method.

\begin{figure}[!ht]
\centering
\includegraphics[width=0.7\linewidth]{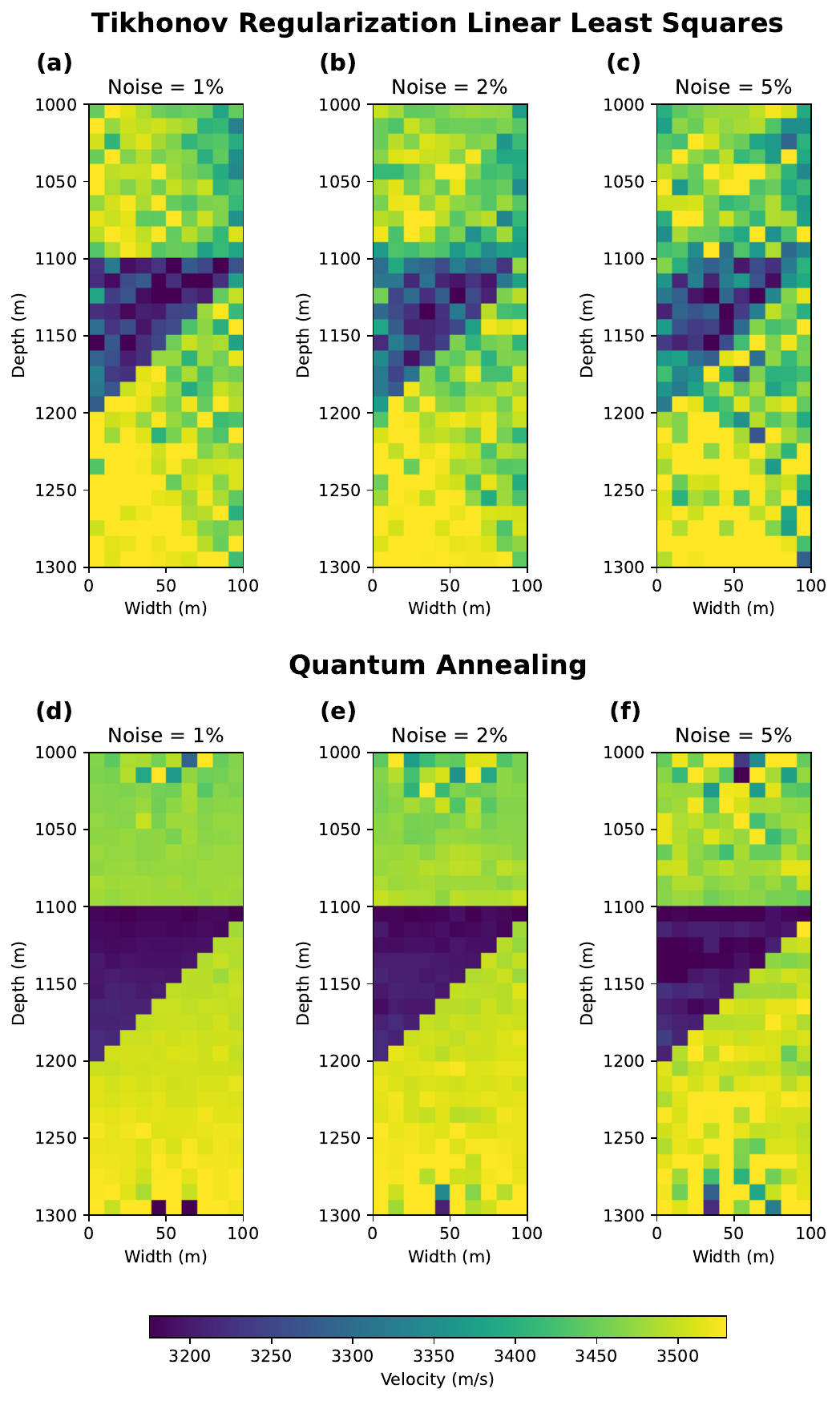}
\caption{Traveltime inversion results with different noisy data using linear least squares (a, b, c) and quantum annealing method (d, e, f)}
\label{fig:combined_velocity_inversions}
\end{figure}

The results reveal a stark contrast in the sensitivity of these methods to noise. In this problem, the Tikhonov regularization linear least squares method is sensitive to noise (Fig. \ref{fig:combined_velocity_inversions}a, b, c). At the noise level of 1\% of the traveltime, while this method can identify the region of carbon storage, the deviation of the inverted model from true model is considerable. As the noise level increases to  2\% and 5\% of the traveltime, the linear least squares method almost fails to accurately determine the velocity model. This significant sensitivity limits its effectiveness in processing noisy seismic data, posing challenges for practical applications.
In contrast, with the same noise, the quantum annealing method is more robust (Fig. \ref{fig:combined_velocity_inversions}d, e, f). At the 1\% noise level, the differences between the noise-free model and the results obtained are small. Differences start to appear primarily in shallow and deep areas where there is less constraint. Remarkably, at the 5\% noise level, the quantum annealing method still effectively reproduces the velocity model. In areas with high ray coverage, these differences are small. This analysis underscores the potential of quantum annealing method for handling noisy seismic data more effectively than the classical linear least squares method.



 

\section*{Discussion}

In this study, we demonstrate several advancements to enhance the robustness and accuracy of seismic traveltime inversion results using the quantum annealing method. A significant improvement involves incorporating non-uniform source and receiver spacing. By using non-uniform spacing, we achieve better ray coverage of both shallow and deep regions, thereby increasing the overall accuracy of the final inverted velocity model.

The initial problem is broken down into smaller sub-problems, we also introduce the boundary \(L\) during the inversion process for better constraint and accommodating hardware limitations. While this paper does not utilize parallel processing, solving these sub-problems independently enables parallel execution, which can significantly reduce computational time.

For practical applications where noise is an unavoidable factor, our method demonstrates a remarkable ability to handle noisy data effectively and is highly suitable for the ill-conditioned problems, maintaining high efficiency and accuracy. The key advantage of the quantum annealing method is its ability to locate the global minimum of the objective function more effectively than classical methods, which are often trapped in local minima. Quantum tunneling allows the quantum system to explore a broader solution space and tunnel barriers that would hinder classical optimization methods \cite{Finnila94}. This method can result in more accurate and reliable inversion outcomes, particularly in complex and heterogeneous environments. The quantum annealer solves problems using principles of quantum mechanics, which inherently depend on the operating environment of the machine. Consequently, running the same problem multiple times may yield slightly different results because of the probabilistic nature of quantum computations \cite{Delgado11, Hevia21}. However, we expect the advancements in quantum technology will decrease this variability and also reduce computational time.

\section*{Conclusion}
The integration of quantum annealing in seismic traveltime inversion represents a potentially major advancement in geophysics. We utilise the open-access D-Wave Advantage quantum annealer to solve the inversion problem for a synthetic velocity model, representing a carbon storage scenario.
The quantum annealing recursive method can give a good inverted velocity result after 10 iterations. Notably, the quantum method outperforms the classical linear least squares method in dealing with noisy data, where classical methods sometimes struggle. Despite promising results, the current state of quantum computing is not without its challenges. The slight variability in the results due to quantum noise \cite{Li_2020, Franca2021, Leon21} requires multiple runs to ensure reliable results. However, these challenges are expected to diminish as advances in quantum technology continue. This study suggests that quantum annealing could revolutionize seismic inversion processes, offering more accurate solutions in scenarios where traditional methods are computationally intensive or even infeasible. As quantum computing matures, its applications in geophysics are likely to expand, encompassing more complex and larger-scale problems. This research not only underscores the potential of quantum annealing in seismic inversion but also sets the stage for future exploration into other areas of inverse problems. We expect that continued development of quantum computing technologies promises to unlock new capabilities, possibly making it an indispensable tool for new challenges.

\section*{Methods}

\subsection*{Quantum Annealing}
Quantum computing is rapidly emerging as a pivotal area of scientific and technological advancement, attracting considerable investment and interest due to its profound potential \cite{britt17, moller17, coccia24}. Unlike classical computers that use bits, which exist only in states of 0 or 1, quantum computers employ quantum bits, or qubits. Qubits possess unique properties such as superposition, entanglement, and interference \cite{qiao18, neeley10, loft20}, enabling them to perform certain complex computations beyond the capabilities of classical computers \cite{Feld19, Neukart17}. Qubits can be constructed from various physical systems such as photons, trapped atoms, nuclear magnetic resonance, quantum dots, dopants in solids, and superconductors \cite{ladd10}. Previous research \cite{baldassi18, Denchev16,albash18, Nakata2014, Senekane21} has provided evidence that quantum computing possibly surpasses classical computers in terms of processing speed and efficiency for certain problems.

The quantum annealing process facilitates the finding of the global minimum of a cost function efficiently. This process can be described using the real-time Schrödinger equation \cite{morita08}:
\begin{equation}
i \hbar \frac{d}{dt} |\Psi(t)\rangle = H(t)|\Psi(t)\rangle
\label{schrodinger}
\end{equation}

where \(| \cdot \rangle \)  is the ket of the Dirac notation \cite{Dirac_1939}, \(i\) is the imaginary unit, \( t \) is time, \( \hbar\) is the reduced Planck's constant, \( \Psi(t) \) is the wave function, \(|\Psi(t)\rangle \) is the quantum state vector, \( H \) is the Hamiltonian representing the total energy of the quantum system \cite{Griffiths_Schroeter_2018, Shankar1994}. If \(\hbar\) is set as 1, the Eq. \ref{schrodinger} becomes:
\begin{equation}
i \frac{d}{dt} |\Psi(t)\rangle = H(t)|\Psi(t)\rangle
\label{schrodingerwithouth}
\end{equation}

The Hamiltonian in quantum annealing can be composed of two components \cite{rajak23, biswas17}:
\begin{equation}
H(t) = A(t) H_0 + B(t) H_1
\end{equation}

where \( H_0 \) is the initial Hamiltonian, representing a system with an initial ground state. \( H_1 \) is the final Hamiltonian, whose ground state encodes the solution to the optimization problem. \( A(t) \) and \( B(t) \) are time-dependent coefficients. \( A(t) \) and \( B(t) \) are set in a range of 0 to 1 so that \( A(t_0) \gg B(t_0) \) at the initial time \( t_0 \) and \( B(t_1) \gg A(t_1) \) at the final time \( t_0 \). During the process, \( A(t) \) monotonically decreases, while \( B(t) \) monotonically increases. At the start of the annealing process, \( H(t) \approx H_0 \). At the end of the annealing process, \( H(t) \approx H_1 \). Thus, the system transitions from the ground state of \( H_0 \) to the ground state of \( H_1 \). If \( H(t) \) changes sufficiently slowly, the state evolves adiabatically \cite{hauke20}.

The problems are then often mapped onto a Quadratic Unconstrained Binary Optimization (QUBO) or Ising model \cite{willsch22}:
\begin{equation}
\text{QUBO:} \quad \min_{x_j = 0,1} \left( \sum_{j \leq k} x_j Q_{jk} x_k + C_1 \right)
\label{eqqubo}
\end{equation}
\begin{equation}
\text{Ising:} \quad \min_{s_j = \pm1} \left( \sum_j h_j s_j + \sum_{j < k} J_{jk} s_j s_k + C_2 \right)
\label{eqising}
\end{equation}
where \(j, k\) are indices, ranging over all qubits. In the QUBO model (Eq. \ref{eqqubo}), \(Q_{jk}\) is the QUBO matrix with values \(Q_{jk} \in \mathbb{R}\). The binary variable vector is \(\mathbf{x}\) with \(x_j \in \{0, 1\}\). In the Ising model (Eq. \ref{eqising}), the problem is defined by the biases \(h_j \in \mathbb{R}\) and the couplers \(J_{jk} \in \mathbb{R}\), and the binary variable vector is \(\mathbf{s}\) with \(s_j \in \{-1, 1\}\). \(C_1\) and \(C_2\) are constants which do not affect the solution of the optimization problem. The Ising model and the QUBO model are mathematically equivalent, allowing them to be translated into each other. This equivalence provides a flexible approach to problem-solving, enabling the conversion of problems between these models based on the requirements and available tools. There are also tools, such as \texttt{ToQUBO.jl}, designed to convert standard problems into the QUBO format \cite{toqubo23}. In this paper, we utilize the quantum annealer from D-Wave Advantage Systems \cite{mcgeoch2020d} to employ direct quantum processing unit (QPU) methods for seismic travel inversion.

\subsection*{Seismic Data Acquisition}

\begin{figure}[!ht]
    \centering
    \includegraphics[width=0.8\linewidth]{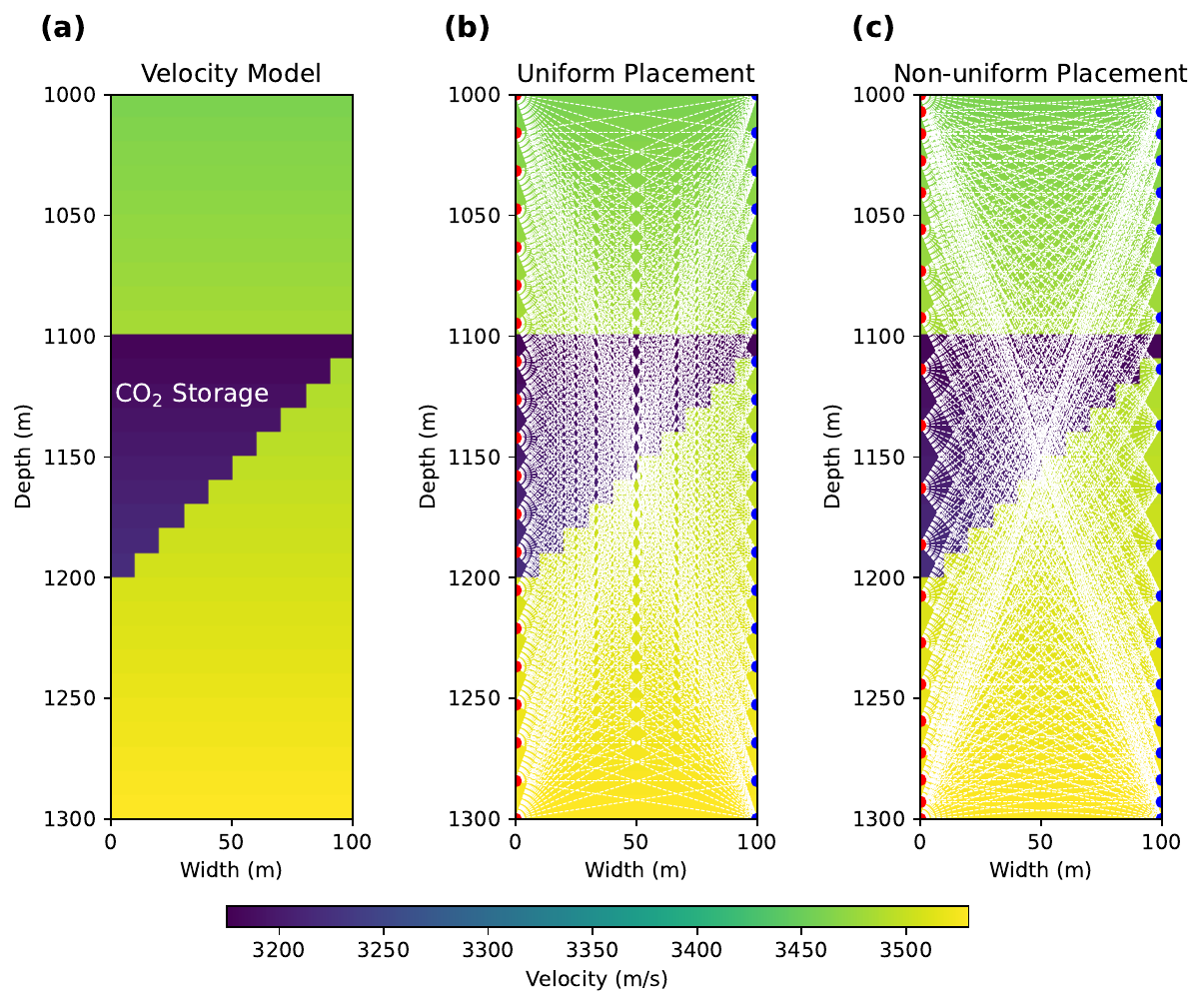}
    \caption{The carbon storage velocity model and ray coverage patterns. Red dots are sources, blue dots are receivers, and white lines represent the ray paths. (a) The synthetic velocity model with a wedge-shaped low-velocity carbon storage formation. (b) Ray coverage from sources and receivers placed in a uniform grid within two boreholes. (c) Ray coverage from sources and receivers placed in a non-uniform pattern, enhancing coverage and constraints for the quantum annealing process.}
    \label{fig:ray_converage}
\end{figure}

We construct a storage velocity model representing carbon storage applications, as shown in Fig. \ref{fig:ray_converage}a. This model spans a depth range from 1000 m to 1300 m and extends 100 m horizontally. The size of the grid cell for this model is 10 x 10 m. Within this model, the carbon storage structure is depicted as a wedge, starting from 1100 m and reaching a maximum depth of 1200 m. The velocity model is constructed with varying velocities to reflect real-world geological conditions. The average velocity within the carbon storage area ranges from 3180 to 3220 m/s, which is about 11\% lower than the surrounding background velocity, which ranges from 3530 to 3640 m/s. Furthermore, the velocities increase with depth.

The uniform placement (Fig. \ref{fig:ray_converage}b) is commonly used in seismic data acquisition for simplicity in boreholes. However, this approach results in significantly lower ray coverage in the shallow and deep sections compared to the middle section. To address these limitations, in our survey, 20 pairs of sources and receivers are non-uniformly deployed within two boreholes (Fig. \ref{fig:ray_converage}c). 
The non-uniform deployment is designed to introduce more constraints for the quantum annealing process, thereby improving the overall accuracy of the seismic inversion. The sources and receivers of non-uniform placement are distributed according to a quadratic polynomial distribution.

\subsection*{Transforming Ray Equations to QUBO}

The set of ray equations can be represented as \cite{mcmechan83}:
\begin{equation}
\mathbf{D} \mathbf{s} = \mathbf{T},
\label{eq1}
\end{equation}
where $\mathbf{D}$ is the matrix of distance increments $d_j$, $\mathbf{s}$ is the slowness vector, and $\mathbf{T}$ is the travel time vector. The size of $\mathbf{D}$ can be very large, therefore solving for $\mathbf{s}$ through matrix operations on $\mathbf{D}$ is computationally intensive. This challenge is exacerbated by the relatively sparse and random distribution of elements within $\mathbf{D}$. Consequently, alternative methods are required to solve these problems efficiently while maintaining accuracy. Quantum annealers can provide quantum metaheuristic algorithms to address this issue. Eq. \ref{eq1} can be solved by minimizing the objective function:
\begin{equation}
f(\mathbf{s}) = \left\| \mathbf{D} \mathbf{s} - \mathbf{T} \right\|^2_2.
\label{eq2}
\end{equation}
The objective function \( f(\mathbf{s}) \) computes the difference between the observed travel times and those predicted by the model given a slowness vector $\mathbf{s}$. Minimizing this function ensures that the model's predictions align as closely as possible with the observed data, thus achieving an optimal fit.
Quantum annealers offer a direct approach to solving binary objective functions \cite{Omalley16}:
\begin{equation}
f(\mathbf{q}) = \left\| \mathbf{A}^d \mathbf{q} - \mathbf{b} \right\|^2_2.
\label{eq3}
\end{equation}
In this formulation, $\mathbf{q}$ is a binary vector, $\mathbf{A}^d$ is a real-valued matrix, and $\mathbf{b}$ is a real-valued vector. Because quantum computers are designed to solve QUBO problems, transforming real-valued variables to binary values is essential. However, the number of binary variables $n_{\text{binary}}$ increases with the number of bits $R$ used for fixed-point approximation, and it is related to the number of real-valued variables $n_{\text{real}}$ as $n_{\text{binary}} = n_{\text{real}} \times R$.
Higher values of $R$ yield greater accuracy in representing the floating-point numbers, but the current limitations of quantum computer hardware restrict the number of qubits available. To address this issue without excessively increasing the number of binary variables, the initial velocity guess, variable boundaries \cite{Souza22} and recursive methods \cite{rogers20} are employed. The initial guess and the boundaries are used to rescale the range of the slowness vector $\mathbf{s}$ to a new vector $\mathbf{x}$ such that $x_i \in [0, 2)$, facilitating easier binary representation. Recursive methods are then applied to enhance the precision of floating-point divisions. These methods iteratively refine the estimate of $\mathbf{s}$, reducing the error at each step.
The initial objective function Eq. \ref{eq2} can be reformulated as:
\begin{equation}
f(\mathbf{x}) = \left\| \mathbf{D} \mathbf{x} - \mathbf{b} \right\|^2_2,
\end{equation}
where $\mathbf{b} = \left(\mathbf{T} + L \mathbf{I} - \mathbf{D} \mathbf{s}_0 \right) / L$, $L$ is the variable boundaries, $\mathbf{s}_0$ is the initial guess of the slowness vector, and $\mathbf{I}$ is the identity vector. The slowness vector $\mathbf{s}$ is then in the range of $[\mathbf{s}_0 - L \mathbf{I}; \mathbf{s}_0 + L \mathbf{I}]$. This range ensures that the solution space is adequately covered. To express this as a binary objective function, $x_i$ is represented in binary form using the $R$ bit fixed-point approximation:
\begin{equation}
x_i = \sum_{r=0}^{R-1} 2^{-r} q_r,
\label{eqxi}
\end{equation}
where $q_r \in \{0, 1\}$ is the value of the $r$-th bit. This transformation is essential for harnessing the computational power of quantum annealers, which are inherently designed to solve binary optimization problems. The new matrix $\mathbf{A}^d$ in Eq. \ref{eq3} is derived from $\mathbf{D}$ and $R$ such that $\mathbf{D} \mathbf{x} = \mathbf{A}^d \mathbf{q}$. The QUBO matrix $Q_{ij}$ in Eq. \ref{eqqubo} is then constructed from the given matrix $\mathbf{D}$ and the calculated vector $\mathbf{b}$ \cite{Omalley16,borle19}:
\begin{align}
Q_{jj} &= \sum_i A_{ij} (A_{ij} - 2b_j) , \\
Q_{jk} &= 2 \sum_i A_{ij} A_{ik} .
\end{align}
The QUBO matrix is directly input into the Quantum Processing Unit (QPU). The system utilizes \texttt{DWaveSampler()} to employ a D-Wave system as the sampler. Subsequently, \texttt{EmbeddingComposite()} manages the mapping between the problem and the D-Wave system’s numerically indexed qubits, a process known as minor-embedding \cite{dwave23}.

In this study, we perform traveltime inversion using \(R=3\) for 10 iterations with quantum annealing. The total number of real-valued variables of the problem is 300. Due to quantum hardware limitations, we break down the model into 30 layers with 10 variables each. This division reduces the complexity of each sub-problem, making it manageable for the quantum processor and allowing for better control of the boundary \(L\). Since the problem from Eq. \ref{eq1} is depth-independent, we simplify the process by adjusting the system's coordinates at each iteration. The approach ensures that each layer is treated independently, reducing the overall complexity of the inversion. By implementing these techniques, we can efficiently solve large-scale traveltime inversion problems using quantum annealers.

\subsection*{Tikhonov Regularization Linear Least Squares Inversion}
For ill-conditioned problems, small changes in \(\mathbf{D}\) or \(\mathbf{T}\) can lead to significant variations in the results \cite{Deif1986}. To mitigate the effects of noise in the data, we employ Tikhonov regularization methods. The new objective function (Eq. \ref{eq2}) can be expressed as a general regularized form \cite{fierro1997}:
\begin{equation}
f_\lambda(\mathbf{s}) = \| \mathbf{D} \mathbf{s} - \mathbf{T} \|_2^2 + \lambda g(\mathbf{s}),
\end{equation}
where \(\lambda\) is the regularization parameter controlling the trade-off between the data fidelity term \(\| \mathbf{D} \mathbf{s} - \mathbf{T} \|_2^2\) and the regularization term \(g(\mathbf{s})\). The Tikhonov regularization is flexible and allows different types of regularization functions. The standard Tikhonov regularization with \( L_2 \)-norm is in the form:
\begin{equation}
f_\lambda(\mathbf{s}) = \| \mathbf{D} \mathbf{s} - \mathbf{T} \|_2^2 + \lambda \|\mathbf{s}\|_2^2
\end{equation}

where \( \|\mathbf{s}\|_2^2 = \mathbf{s}^T \mathbf{s} \) penalizes large values in the solution. Another form is the first-order Tikhonov regularization with a smoothness regularization:
\begin{equation}
f_\lambda(\mathbf{s}) = \| \mathbf{D} \mathbf{s} - \mathbf{T} \|_2^2 + \lambda \| D_1 \mathbf{s} \|_2^2
\end{equation}

where \( D_1 \) is the first-order difference operator which enforces smooth variation in \( \mathbf{s} \) by penalizing large first derivatives. Similarly, second-order Tikhonov regularization penalizes the curvature of the solution:
\begin{equation}
f_\lambda(\mathbf{s}) = \| \mathbf{D} \mathbf{s} - \mathbf{T} \|_2^2 + \lambda \| D_2 \mathbf{s} \|_2^2
\end{equation}

where \( D_2 \) is the second-order difference operator which enforces smooth curvature by penalizing large second derivatives.
In general, \( g(\mathbf{s}) \) can be \( g(\mathbf{s}) = \|\mathbf{s}\|_2^2 \) for standard \( L_2 \)-norm regularization, \( g(\mathbf{s}) = \|D_1 \mathbf{s} \|_2^2 \) for first-order smoothness, or \( g(\mathbf{s}) = \|D_2 \mathbf{s} \|_2^2 \) for second-order smoothness. The choice of \( g(\mathbf{s}) \) depends on prior knowledge and the desired properties of the solution. Here, we use a custom Tikhonov regularization \(g(\mathbf{s}) = \| \mathbf{s} - \mathbf{s}_0 \|_2^2\), where \(\mathbf{s}_0\) is the initial guess for the slowness and is chosen as the input for the quantum annealing process. The objective function is now expressed as:
\begin{equation}
f_\lambda(\mathbf{s}) = \| \mathbf{D} \mathbf{s} - \mathbf{T} \|_2^2 + \lambda \| \mathbf{s} - \mathbf{s}_0 \|_2^2.
\end{equation}
The solution to this regularized problem is given by:
\begin{equation}
\mathbf{s} = \left( \mathbf{D}^T \mathbf{D} + \lambda \mathbf{I} \right)^{-1} \left( \mathbf{D}^T \mathbf{T} + \lambda \mathbf{s}_0 \right).
\label{tik solution}
\end{equation}

\subsection*{Cost Analysis}

Quantum annealing directly solves \(f(\mathbf{q}) = \left\| \mathbf{A}^d \mathbf{q} - \mathbf{b} \right\|^2_2\) (Eq. \ref{eq3}) by providing the solution binary vector \(\mathbf{q}\). The quantum annealing process involves three sources of computational cost: preparing the binary problem, executing the annealing on the quantum hardware, and post-processing the results. The cost of preparing and post-processing is \(\mathcal{O}(mn^2 + mnc + n^2c^2)\). For a single iteration of the loop, the cost of executing the annealing process on the quantum hardware is \(\mathcal{O}(\mathcal{P}(cn))\). Here, \(m\) and \(n\) represent the rows and columns of the matrix \(\mathbf{D}\), respectively. \(\mathcal{P}(cn)\) is a polynomial term. The parameter \(c = |\Theta| + 1\), where \(\Theta\) used for representing the variables \(x_j\) as a fixed-point approximation in terms of power of 2 (Eq. \ref{eqxi}). The general form of the set \(\Theta\) is defined as \cite{borle19}:
\begin{equation}
    \Theta = \{ 2^l : l \in [o, p] \land l, o, p \in \mathbb{Z} \},
\end{equation}
where \(l\) is contiguous integer values within the interval \([o, p]\), and \(o\) and \(p\) are the lower and upper bounds of the interval. In our scheme, instead of using a large range for \(o\) and \(p\), we refine the precision of \(x_i\) iteratively over multiple loops to achieve higher accuracy. For a 3-bit fixed-point approximation with $x_i \in [0, 2)$, \(o = 0\) and \(p = -2\). For the \(K\) iterations, the total cost for  process is:
\begin{equation}
    \mathcal{O}(mn^2 + mnc + n^2c^2 + K\cdot\mathcal{P}(cn)),
    \label{cost anl}
\end{equation}

For Tikhonov regularization methods with the direct solver (e.g., LU decomposition), the computational cost primarily depends on solving the modified linear system with the solution presented in Eq.~\ref{tik solution}. The cost for finding \(\mathbf{s}\) with a given \(\lambda\) is \(\mathcal{O}(mn^2 + n^3)\). Selecting the optimal \(\lambda\) is crucial for achieving a balance between data fidelity and regularization. One commonly used technique is the L-curve method, which plots the norm of the regularization term \(\|R(\mathbf{s})\|\) against the residual norm \(\|\mathbf{D} \mathbf{s} - \mathbf{T}\|\) on a log-log scale \cite{Hansen2000, CALVETTI2000423}. For \(N_\lambda\) values of \(\lambda\), the cost of solving the Tikhonov-regularized system for each \(\lambda\) is \( \mathcal{O}(N_\lambda \cdot (mn^2 + n^3))\). In addition, the computation of the residual norm \(\|\mathbf{D} \mathbf{s} - \mathbf{T}\|\) for each \(\lambda\) involves matrix-vector multiplications, contributing a cost of \(\mathcal{O}(N_\lambda \cdot mn)\). If we want to find the optimal \(\lambda\) automatically, the curvature of the L-curve must be computed. This requires additional operations such as numerical differentiation and curvature estimation, which are negligible compared to the cost of solving the regularized systems. Thus, the total cost for the L-curve method, including the automatic selection of \(\lambda\), is:
\begin{equation}
    \mathcal{O}(N_\lambda \cdot (mn^2 + n^3)).
    \label{cost tik}
\end{equation}

From Eqs. \ref{cost anl} and \ref{cost tik}, achieving a computational cost of \(\mathcal{O}(\mathcal{P}(cn)) < \mathcal{O}(n^3)\) would enable quantum annealing to significantly accelerate problem-solving. Research efforts, such as those utilizing multi-qubit correction techniques \cite{dorband2018methodfindinglowerenergy}, aim to realize this improvement. These approaches can potentially achieve a substantial reduction in computational complexity. This advancement would facilitate rapid convergence to the global minimum and unlock considerable performance gains.

\bibliography{references}

\begin{thebibliography}{10}
\urlstyle{rm}
\expandafter\ifx\csname url\endcsname\relax
  \def\url#1{\texttt{#1}}\fi
\expandafter\ifx\csname urlprefix\endcsname\relax\def\urlprefix{URL }\fi
\expandafter\ifx\csname doiprefix\endcsname\relax\def\doiprefix{DOI: }\fi
\providecommand{\bibinfo}[2]{#2}
\providecommand{\eprint}[2][]{\url{#2}}

\bibitem{moradi18}
\bibinfo{author}{Moradi, S.}, \bibinfo{author}{Trad, D.} \& \bibinfo{author}{Innanen, K.~A.}
\newblock \emph{\bibinfo{title}{Quantum computing in geophysics: Algorithms, computational costs, and future applications}}, \bibinfo{pages}{4649--4653} (\bibinfo{publisher}{Society of Exploration Geophysicists}, \bibinfo{year}{2018}).
\newblock \eprint{https://library.seg.org/doi/pdf/10.1190/segam2018-2998507.1}.

\bibitem{sarkar18}
\bibinfo{author}{Sarkar, R.} \& \bibinfo{author}{Levin, S.~A.}
\newblock \emph{\bibinfo{title}{Snell tomography for net-to-gross estimation using quantum annealing}}, \bibinfo{pages}{5078--5082} (\bibinfo{publisher}{Society of Exploration Geophysicists}, \bibinfo{year}{2018}).
\newblock \eprint{https://library.seg.org/doi/pdf/10.1190/segam2018-2998409.1}.

\bibitem{greer20}
\bibinfo{author}{Greer, S.} \& \bibinfo{author}{O’Malley, D.}
\newblock \emph{\bibinfo{title}{An approach to seismic inversion with quantum annealing}}, \bibinfo{pages}{2845--2849} (\bibinfo{publisher}{Society of Exploration Geophysicists}, \bibinfo{year}{2020}).
\newblock \eprint{https://library.seg.org/doi/pdf/10.1190/segam2020-3424413.1}.

\bibitem{kumar22}
\bibinfo{editor}{of~Petroleum~Engineers, S.} (ed.).
\newblock \emph{\bibinfo{title}{{Quantum Computation for End-to-End Seismic Data Processing with Its Computational Advantages and Economic Sustainability}}}, vol. \bibinfo{volume}{Day 2 Tue, November 01, 2022} of \emph{\bibinfo{series}{Abu Dhabi International Petroleum Exhibition and Conference}} (\bibinfo{year}{2022}).
\newblock \doiprefix\url{10.2118/211843-MS}.
\newblock \eprint{https://onepetro.org/SPEADIP/proceedings-pdf/22ADIP/2-22ADIP/D021S037R002/3043678/spe-211843-ms.pdf}.

\bibitem{wang24}
\bibinfo{author}{Wang, S.}, \bibinfo{author}{Liu, C.}, \bibinfo{author}{Li, P.}, \bibinfo{author}{Chen, C.} \& \bibinfo{author}{Song, C.}
\newblock \bibinfo{journal}{\bibinfo{title}{{Stable and efficient seismic impedance inversion using quantum annealing with L1 norm regularization}}}.
\newblock {\emph{\JournalTitle{Journal of Geophysics and Engineering}}} \textbf{\bibinfo{volume}{21}}, \bibinfo{pages}{330--343}, \doiprefix\url{10.1093/jge/gxae003} (\bibinfo{year}{2024}).
\newblock \eprint{https://academic.oup.com/jge/article-pdf/21/1/330/56746369/gxae003.pdf}.

\bibitem{Yulianti22}
\bibinfo{author}{Yulianti, L.~P.} \& \bibinfo{author}{Surendro, K.}
\newblock \bibinfo{journal}{\bibinfo{title}{Implementation of quantum annealing: A systematic review}}.
\newblock {\emph{\JournalTitle{IEEE Access}}} \textbf{\bibinfo{volume}{10}}, \bibinfo{pages}{73156--73177}, \doiprefix\url{10.1109/ACCESS.2022.3188117} (\bibinfo{year}{2022}).

\bibitem{MCGEOCH20}
\bibinfo{author}{McGeoch, C.}
\newblock \bibinfo{journal}{\bibinfo{title}{Theory versus practice in annealing-based quantum computing}}.
\newblock {\emph{\JournalTitle{Theoretical Computer Science}}} \textbf{\bibinfo{volume}{816}}, \bibinfo{pages}{169--183}, \doiprefix\url{https://doi.org/10.1016/j.tcs.2020.01.024} (\bibinfo{year}{2020}).

\bibitem{crosson16}
\bibinfo{author}{Crosson, E.} \& \bibinfo{author}{Harrow, A.~W.}
\newblock \bibinfo{title}{Simulated quantum annealing can be exponentially faster than classical simulated annealing}.
\newblock In \emph{\bibinfo{booktitle}{2016 IEEE 57th Annual Symposium on Foundations of Computer Science (FOCS)}}, \bibinfo{pages}{714--723}, \doiprefix\url{10.1109/FOCS.2016.81} (\bibinfo{year}{2016}).

\bibitem{nimbe21}
\bibinfo{author}{Nimbe, P.}, \bibinfo{author}{Weyori, B.~A.} \& \bibinfo{author}{Adekoya, A.~F.}
\newblock \bibinfo{journal}{\bibinfo{title}{Models in quantum computing: a systematic review}}.
\newblock {\emph{\JournalTitle{Quantum Information Processing}}} \textbf{\bibinfo{volume}{20}}, \bibinfo{pages}{80}, \doiprefix\url{10.1007/s11128-021-03021-3} (\bibinfo{year}{2021}).

\bibitem{Lu2023}
\bibinfo{author}{Lu, B.}, \bibinfo{author}{Liu, L.}, \bibinfo{author}{Song, J.-Y.}, \bibinfo{author}{Wen, K.} \& \bibinfo{author}{Wang, C.}
\newblock \bibinfo{journal}{\bibinfo{title}{Recent progress on coherent computation based on quantum squeezing}}.
\newblock {\emph{\JournalTitle{AAPPS Bulletin}}} \textbf{\bibinfo{volume}{33}}, \bibinfo{pages}{7}, \doiprefix\url{10.1007/s43673-023-00077-4} (\bibinfo{year}{2023}).

\bibitem{Alulaiw15}
\emph{\bibinfo{title}{{Prestack Seismic Inversion by Quantum Annealing: Application To Cana Field}}}, vol. \bibinfo{volume}{All Days} of \emph{\bibinfo{series}{SEG International Exposition and Annual Meeting}}.
\newblock \eprint{https://onepetro.org/SEGAM/proceedings-pdf/SEG15/All-SEG15/SEG-2015-5831164/1470492/seg-2015-5831164.pdf}.

\bibitem{Albino22}
\bibinfo{author}{Albino, A.~S.}, \bibinfo{author}{Pires, O.~M.}, \bibinfo{author}{Nogueira, P.}, \bibinfo{author}{de~Souza, R.~F.} \& \bibinfo{author}{Nascimento, E. G.~S.}
\newblock \bibinfo{title}{Quantum computational intelligence for traveltime seismic inversion} (\bibinfo{year}{2022}).
\newblock \eprint{2208.05794}.

\bibitem{mcmechan83}
\bibinfo{author}{McMechan, G.~A.}
\newblock \bibinfo{journal}{\bibinfo{title}{{Seismic tomography in boreholes}}}.
\newblock {\emph{\JournalTitle{Geophysical Journal International}}} \textbf{\bibinfo{volume}{74}}, \bibinfo{pages}{601--612}, \doiprefix\url{10.1111/j.1365-246X.1983.tb01891.x} (\bibinfo{year}{1983}).
\newblock \eprint{https://academic.oup.com/gji/article-pdf/74/2/601/1757693/74-2-601.pdf}.

\bibitem{mcgeoch2020d}
\bibinfo{author}{McGeoch, C.} \& \bibinfo{author}{Farr{\'e}, P.}
\newblock \bibinfo{journal}{\bibinfo{title}{The d-wave advantage system: An overview}}.
\newblock {\emph{\JournalTitle{D-Wave Systems Inc., Burnaby, BC, Canada, Tech. Rep}}}  (\bibinfo{year}{2020}).

\bibitem{Souza22}
\bibinfo{author}{Souza, A.~M.} \emph{et~al.}
\newblock \bibinfo{journal}{\bibinfo{title}{An application of quantum annealing computing to seismic inversion}}.
\newblock {\emph{\JournalTitle{Frontiers in Physics}}} \textbf{\bibinfo{volume}{9}}, \doiprefix\url{10.3389/fphy.2021.748285} (\bibinfo{year}{2022}).

\bibitem{Daley2008}
\bibinfo{author}{Daley, T.~M.}, \bibinfo{author}{Myer, L.~R.}, \bibinfo{author}{Peterson, J.~E.}, \bibinfo{author}{Majer, E.~L.} \& \bibinfo{author}{Hoversten, G.~M.}
\newblock \bibinfo{journal}{\bibinfo{title}{Time-lapse crosswell seismic and vsp monitoring of injected co2 in a brine aquifer}}.
\newblock {\emph{\JournalTitle{Environmental Geology}}} \textbf{\bibinfo{volume}{54}}, \bibinfo{pages}{1657--1665}, \doiprefix\url{10.1007/s00254-007-0943-z} (\bibinfo{year}{2008}).

\bibitem{Finnila94}
\bibinfo{author}{Finnila, A.}, \bibinfo{author}{Gomez, M.}, \bibinfo{author}{Sebenik, C.}, \bibinfo{author}{Stenson, C.} \& \bibinfo{author}{Doll, J.}
\newblock \bibinfo{journal}{\bibinfo{title}{Quantum annealing: A new method for minimizing multidimensional functions}}.
\newblock {\emph{\JournalTitle{Chemical Physics Letters}}} \textbf{\bibinfo{volume}{219}}, \bibinfo{pages}{343--348}, \doiprefix\url{https://doi.org/10.1016/0009-2614(94)00117-0} (\bibinfo{year}{1994}).

\bibitem{Delgado11}
\bibinfo{author}{P\'erez-Delgado, C.~A.} \& \bibinfo{author}{Kok, P.}
\newblock \bibinfo{journal}{\bibinfo{title}{Quantum computers: Definition and implementations}}.
\newblock {\emph{\JournalTitle{Phys. Rev. A}}} \textbf{\bibinfo{volume}{83}}, \bibinfo{pages}{012303}, \doiprefix\url{10.1103/PhysRevA.83.012303} (\bibinfo{year}{2011}).

\bibitem{Hevia21}
\bibinfo{author}{Hevia, J.~L.}, \bibinfo{author}{Peterssen, G.}, \bibinfo{author}{Ebert, C.} \& \bibinfo{author}{Piattini, M.}
\newblock \bibinfo{journal}{\bibinfo{title}{Quantum computing}}.
\newblock {\emph{\JournalTitle{IEEE Software}}} \textbf{\bibinfo{volume}{38}}, \bibinfo{pages}{7--15}, \doiprefix\url{10.1109/MS.2021.3087755} (\bibinfo{year}{2021}).

\bibitem{Li_2020}
\bibinfo{author}{Li, R.~Y.}, \bibinfo{author}{Albash, T.} \& \bibinfo{author}{Lidar, D.~A.}
\newblock \bibinfo{journal}{\bibinfo{title}{Limitations of error corrected quantum annealing in improving the performance of boltzmann machines}}.
\newblock {\emph{\JournalTitle{Quantum Science and Technology}}} \textbf{\bibinfo{volume}{5}}, \bibinfo{pages}{045010}, \doiprefix\url{10.1088/2058-9565/ab9aab} (\bibinfo{year}{2020}).

\bibitem{Franca2021}
\bibinfo{author}{França, D.~S.} \& \bibinfo{author}{García-Patrón, R.}
\newblock \bibinfo{journal}{\bibinfo{title}{Limitations of optimization algorithms on noisy quantum devices}}.
\newblock {\emph{\JournalTitle{Nature Physics}}} \textbf{\bibinfo{volume}{17}}, \bibinfo{pages}{1221--1227}, \doiprefix\url{10.1038/s41567-021-01356-3} (\bibinfo{year}{2021}).

\bibitem{Leon21}
\bibinfo{author}{de~Leon, N.~P.} \emph{et~al.}
\newblock \bibinfo{journal}{\bibinfo{title}{Materials challenges and opportunities for quantum computing hardware}}.
\newblock {\emph{\JournalTitle{Science}}} \textbf{\bibinfo{volume}{372}}, \bibinfo{pages}{eabb2823}, \doiprefix\url{10.1126/science.abb2823} (\bibinfo{year}{2021}).
\newblock \eprint{https://www.science.org/doi/pdf/10.1126/science.abb2823}.

\bibitem{britt17}
\bibinfo{author}{Britt, K.~A.} \& \bibinfo{author}{Humble, T.~S.}
\newblock \bibinfo{journal}{\bibinfo{title}{High-performance computing with quantum processing units}}.
\newblock {\emph{\JournalTitle{J. Emerg. Technol. Comput. Syst.}}} \textbf{\bibinfo{volume}{13}}, \doiprefix\url{10.1145/3007651} (\bibinfo{year}{2017}).

\bibitem{moller17}
\bibinfo{author}{Möller, M.} \& \bibinfo{author}{Vuik, C.}
\newblock \bibinfo{journal}{\bibinfo{title}{On the impact of quantum computing technology on future developments in high-performance scientific computing}}.
\newblock {\emph{\JournalTitle{Ethics and Information Technology}}} \textbf{\bibinfo{volume}{19}}, \bibinfo{pages}{253--269}, \doiprefix\url{10.1007/s10676-017-9438-0} (\bibinfo{year}{2017}).

\bibitem{coccia24}
\bibinfo{author}{Coccia, M.}, \bibinfo{author}{Roshani, S.} \& \bibinfo{author}{Mosleh, M.}
\newblock \bibinfo{journal}{\bibinfo{title}{Evolution of quantum computing: Theoretical and innovation management implications for emerging quantum industry}}.
\newblock {\emph{\JournalTitle{IEEE Transactions on Engineering Management}}} \textbf{\bibinfo{volume}{71}}, \bibinfo{pages}{2270--2280}, \doiprefix\url{10.1109/TEM.2022.3175633} (\bibinfo{year}{2024}).

\bibitem{qiao18}
\bibinfo{author}{Qiao, L.-F.} \emph{et~al.}
\newblock \bibinfo{journal}{\bibinfo{title}{Entanglement activation from quantum coherence and superposition}}.
\newblock {\emph{\JournalTitle{Phys. Rev. A}}} \textbf{\bibinfo{volume}{98}}, \bibinfo{pages}{052351}, \doiprefix\url{10.1103/PhysRevA.98.052351} (\bibinfo{year}{2018}).

\bibitem{neeley10}
\bibinfo{author}{Neeley, M.} \emph{et~al.}
\newblock \bibinfo{journal}{\bibinfo{title}{Generation of three-qubit entangled states using superconducting phase qubits}}.
\newblock {\emph{\JournalTitle{Nature}}} \textbf{\bibinfo{volume}{467}}, \bibinfo{pages}{570--573}, \doiprefix\url{10.1038/nature09418} (\bibinfo{year}{2010}).

\bibitem{loft20}
\bibinfo{author}{Loft, N. J.~S.} \emph{et~al.}
\newblock \bibinfo{journal}{\bibinfo{title}{Quantum interference device for controlled two-qubit operations}}.
\newblock {\emph{\JournalTitle{npj Quantum Information}}} \textbf{\bibinfo{volume}{6}}, \bibinfo{pages}{47}, \doiprefix\url{10.1038/s41534-020-0275-3} (\bibinfo{year}{2020}).

\bibitem{Feld19}
\bibinfo{author}{Feld, S.} \emph{et~al.}
\newblock \bibinfo{journal}{\bibinfo{title}{A hybrid solution method for the capacitated vehicle routing problem using a quantum annealer}}.
\newblock {\emph{\JournalTitle{Frontiers in ICT}}} \textbf{\bibinfo{volume}{6}}, \doiprefix\url{10.3389/fict.2019.00013} (\bibinfo{year}{2019}).

\bibitem{Neukart17}
\bibinfo{author}{Neukart, F.} \emph{et~al.}
\newblock \bibinfo{journal}{\bibinfo{title}{Traffic flow optimization using a quantum annealer}}.
\newblock {\emph{\JournalTitle{Frontiers in ICT}}} \textbf{\bibinfo{volume}{4}}, \doiprefix\url{10.3389/fict.2017.00029} (\bibinfo{year}{2017}).

\bibitem{ladd10}
\bibinfo{author}{Ladd, T.~D.} \emph{et~al.}
\newblock \bibinfo{journal}{\bibinfo{title}{Quantum computers}}.
\newblock {\emph{\JournalTitle{Nature}}} \textbf{\bibinfo{volume}{464}}, \bibinfo{pages}{45--53}, \doiprefix\url{10.1038/nature08812} (\bibinfo{year}{2010}).

\bibitem{baldassi18}
\bibinfo{author}{Baldassi, C.} \& \bibinfo{author}{Zecchina, R.}
\newblock \bibinfo{journal}{\bibinfo{title}{Efficiency of quantum vs. classical annealing in nonconvex learning problems}}.
\newblock {\emph{\JournalTitle{Proceedings of the National Academy of Sciences}}} \textbf{\bibinfo{volume}{115}}, \bibinfo{pages}{1457--1462}, \doiprefix\url{10.1073/pnas.1711456115} (\bibinfo{year}{2018}).
\newblock \eprint{https://www.pnas.org/doi/pdf/10.1073/pnas.1711456115}.

\bibitem{Denchev16}
\bibinfo{author}{Denchev, V.~S.} \emph{et~al.}
\newblock \bibinfo{journal}{\bibinfo{title}{What is the computational value of finite-range tunneling?}}
\newblock {\emph{\JournalTitle{Phys. Rev. X}}} \textbf{\bibinfo{volume}{6}}, \bibinfo{pages}{031015}, \doiprefix\url{10.1103/PhysRevX.6.031015} (\bibinfo{year}{2016}).

\bibitem{albash18}
\bibinfo{author}{Albash, T.} \& \bibinfo{author}{Lidar, D.~A.}
\newblock \bibinfo{journal}{\bibinfo{title}{Demonstration of a scaling advantage for a quantum annealer over simulated annealing}}.
\newblock {\emph{\JournalTitle{Phys. Rev. X}}} \textbf{\bibinfo{volume}{8}}, \bibinfo{pages}{031016}, \doiprefix\url{10.1103/PhysRevX.8.031016} (\bibinfo{year}{2018}).

\bibitem{Nakata2014}
\bibinfo{author}{Nakata, Y.} \& \bibinfo{author}{Murao, M.}
\newblock \bibinfo{journal}{\bibinfo{title}{Diagonal quantum circuits: Their computational power and applications}}.
\newblock {\emph{\JournalTitle{The European Physical Journal Plus}}} \textbf{\bibinfo{volume}{129}}, \bibinfo{pages}{152}, \doiprefix\url{10.1140/epjp/i2014-14152-9} (\bibinfo{year}{2014}).

\bibitem{Senekane21}
\bibinfo{author}{Senekane, M.}
\newblock \emph{\bibinfo{title}{Hands-on quantum information processing with Python : get up and running with information processing and computing based on quantum mechanics using Python}} (\bibinfo{publisher}{Packt}, \bibinfo{address}{Birmingham}, \bibinfo{year}{2021}).
\newblock \bibinfo{note}{Englisch}.

\bibitem{morita08}
\bibinfo{author}{Morita, S.} \& \bibinfo{author}{Nishimori, H.}
\newblock \bibinfo{journal}{\bibinfo{title}{{Mathematical foundation of quantum annealing}}}.
\newblock {\emph{\JournalTitle{Journal of Mathematical Physics}}} \textbf{\bibinfo{volume}{49}}, \bibinfo{pages}{125210}, \doiprefix\url{10.1063/1.2995837} (\bibinfo{year}{2008}).
\newblock \eprint{https://pubs.aip.org/aip/jmp/article-pdf/doi/10.1063/1.2995837/13869474/125210\_1\_online.pdf}.

\bibitem{Dirac_1939}
\bibinfo{author}{Dirac, P. A.~M.}
\newblock \bibinfo{journal}{\bibinfo{title}{A new notation for quantum mechanics}}.
\newblock {\emph{\JournalTitle{Mathematical Proceedings of the Cambridge Philosophical Society}}} \textbf{\bibinfo{volume}{35}}, \bibinfo{pages}{416–418}, \doiprefix\url{10.1017/S0305004100021162} (\bibinfo{year}{1939}).

\bibitem{Griffiths_Schroeter_2018}
\bibinfo{author}{Griffiths, D.~J.} \& \bibinfo{author}{Schroeter, D.~F.}
\newblock \emph{\bibinfo{title}{Introduction to Quantum Mechanics}} (\bibinfo{publisher}{Cambridge University Press}, \bibinfo{year}{2018}), \bibinfo{edition}{3} edn.

\bibitem{Shankar1994}
\bibinfo{author}{Shankar, R.}
\newblock \emph{\bibinfo{title}{The Postulates---a General Discussion}}, \bibinfo{pages}{115--150} (\bibinfo{publisher}{Springer US}, \bibinfo{address}{New York, NY}, \bibinfo{year}{1994}).

\bibitem{rajak23}
\bibinfo{author}{Rajak, A.}, \bibinfo{author}{Suzuki, S.}, \bibinfo{author}{Dutta, A.} \& \bibinfo{author}{Chakrabarti, B.~K.}
\newblock \bibinfo{journal}{\bibinfo{title}{Quantum annealing: an overview}}.
\newblock {\emph{\JournalTitle{Philosophical Transactions of the Royal Society A: Mathematical, Physical and Engineering Sciences}}} \textbf{\bibinfo{volume}{381}}, \bibinfo{pages}{20210417}, \doiprefix\url{10.1098/rsta.2021.0417} (\bibinfo{year}{2023}).
\newblock \eprint{https://royalsocietypublishing.org/doi/pdf/10.1098/rsta.2021.0417}.

\bibitem{biswas17}
\bibinfo{author}{Biswas, R.} \emph{et~al.}
\newblock \bibinfo{journal}{\bibinfo{title}{A nasa perspective on quantum computing: Opportunities and challenges}}.
\newblock {\emph{\JournalTitle{Parallel Computing}}} \textbf{\bibinfo{volume}{64}}, \bibinfo{pages}{81--98}, \doiprefix\url{10.1016/j.parco.2016.11.002} (\bibinfo{year}{2017}).
\newblock \bibinfo{note}{High-End Computing for Next-Generation Scientific Discovery}.

\bibitem{hauke20}
\bibinfo{author}{Hauke, P.}, \bibinfo{author}{Katzgraber, H.~G.}, \bibinfo{author}{Lechner, W.}, \bibinfo{author}{Nishimori, H.} \& \bibinfo{author}{Oliver, W.~D.}
\newblock \bibinfo{journal}{\bibinfo{title}{Perspectives of quantum annealing: methods and implementations}}.
\newblock {\emph{\JournalTitle{Reports on Progress in Physics}}} \textbf{\bibinfo{volume}{83}}, \bibinfo{pages}{054401}, \doiprefix\url{10.1088/1361-6633/ab85b8} (\bibinfo{year}{2020}).

\bibitem{willsch22}
\bibinfo{author}{Willsch, D.} \emph{et~al.}
\newblock \bibinfo{journal}{\bibinfo{title}{Benchmarking advantage and d-wave 2000q quantum annealers with exact cover problems}}.
\newblock {\emph{\JournalTitle{Quantum Information Processing}}} \textbf{\bibinfo{volume}{21}}, \bibinfo{pages}{141}, \doiprefix\url{10.1007/s11128-022-03476-y} (\bibinfo{year}{2022}).

\bibitem{toqubo23}
\bibinfo{author}{Xavier, P.~M.}, \bibinfo{author}{Ripper, P.}, \bibinfo{author}{Andrade, T.}, \bibinfo{author}{Garcia, J.~D.} \& \bibinfo{author}{Neira, D. E.~B.}
\newblock \bibinfo{title}{{ToQUBO.jl}}, \doiprefix\url{10.5281/zenodo.7644291} (\bibinfo{year}{2023}).

\bibitem{Omalley16}
\bibinfo{author}{O'Malley, D.} \& \bibinfo{author}{Vesselinov, V.~V.}
\newblock \bibinfo{title}{Toq.jl: A high-level programming language for d-wave machines based on julia}.
\newblock In \emph{\bibinfo{booktitle}{2016 IEEE High Performance Extreme Computing Conference (HPEC)}}, \bibinfo{pages}{1--7}, \doiprefix\url{10.1109/HPEC.2016.7761616} (\bibinfo{year}{2016}).

\bibitem{rogers20}
\bibinfo{author}{Rogers, M.~L.} \& \bibinfo{author}{Singleton, R.~L.}
\newblock \bibinfo{journal}{\bibinfo{title}{Floating-point calculations on a quantum annealer: Division and matrix inversion}}.
\newblock {\emph{\JournalTitle{Frontiers in Physics}}} \textbf{\bibinfo{volume}{8}}, \doiprefix\url{10.3389/fphy.2020.00265} (\bibinfo{year}{2020}).

\bibitem{borle19}
\bibinfo{author}{Borle, A.} \& \bibinfo{author}{Lomonaco, S.~J.}
\newblock \bibinfo{title}{Analyzing the quantum annealing approach for solving linear least squares problems}.
\newblock In \bibinfo{editor}{Das, G.~K.}, \bibinfo{editor}{Mandal, P.~S.}, \bibinfo{editor}{Mukhopadhyaya, K.} \& \bibinfo{editor}{Nakano, S.-i.} (eds.) \emph{\bibinfo{booktitle}{WALCOM: Algorithms and Computation}}, \bibinfo{pages}{289--301} (\bibinfo{publisher}{Springer International Publishing}, \bibinfo{address}{Cham}, \bibinfo{year}{2019}).

\bibitem{dwave23}
\bibinfo{author}{{D-Wave Systems Inc}}.
\newblock \emph{\bibinfo{title}{dwave-system Documentation}} (\bibinfo{year}{2023}).
\newblock \bibinfo{note}{Release 1.18.0}.

\bibitem{Deif1986}
\bibinfo{author}{Deif, A.}
\newblock \emph{\bibinfo{title}{Perturbation of Linear Equations}}, \bibinfo{pages}{1--43} (\bibinfo{publisher}{Springer Berlin Heidelberg}, \bibinfo{address}{Berlin, Heidelberg}, \bibinfo{year}{1986}).

\bibitem{fierro1997}
\bibinfo{author}{Fierro, R.~D.}, \bibinfo{author}{Golub, G.~H.}, \bibinfo{author}{Hansen, P.~C.} \& \bibinfo{author}{O'Leary, D.~P.}
\newblock \bibinfo{journal}{\bibinfo{title}{Regularization by truncated total least squares}}.
\newblock {\emph{\JournalTitle{SIAM Journal on Scientific Computing}}} \textbf{\bibinfo{volume}{18}}, \bibinfo{pages}{1223--1241}, \doiprefix\url{10.1137/S1064827594263837} (\bibinfo{year}{1997}).
\newblock \eprint{https://doi.org/10.1137/S1064827594263837}.

\bibitem{Hansen2000}
\bibinfo{author}{Hansen, P.}
\newblock \bibinfo{title}{The l-curve and its use in the numerical treatment of inverse problems}.
\newblock In \emph{\bibinfo{booktitle}{InviteComputational Inverse Problems in Electrocardiology}} (\bibinfo{publisher}{WIT Press}, \bibinfo{year}{2000}).
\newblock \bibinfo{note}{InviteComputational Inverse Problems in Electrocardiology ; Conference date: 01-01-2000}.

\bibitem{CALVETTI2000423}
\bibinfo{author}{Calvetti, D.}, \bibinfo{author}{Morigi, S.}, \bibinfo{author}{Reichel, L.} \& \bibinfo{author}{Sgallari, F.}
\newblock \bibinfo{journal}{\bibinfo{title}{Tikhonov regularization and the l-curve for large discrete ill-posed problems}}.
\newblock {\emph{\JournalTitle{Journal of Computational and Applied Mathematics}}} \textbf{\bibinfo{volume}{123}}, \bibinfo{pages}{423--446}, \doiprefix\url{https://doi.org/10.1016/S0377-0427(00)00414-3} (\bibinfo{year}{2000}).
\newblock \bibinfo{note}{Numerical Analysis 2000. Vol. III: Linear Algebra}.

\bibitem{dorband2018methodfindinglowerenergy}
\bibinfo{author}{Dorband, J.~E.}
\newblock \bibinfo{title}{A method of finding a lower energy solution to a qubo/ising objective function} (\bibinfo{year}{2018}).
\newblock \eprint{1801.04849}.

\end{thebibliography}

\section*{Acknowledgements}

We would like to thank the Reservoir Characterization Project (RCP), Colorado School of Mines, for the financial support.

\section*{Author contributions statement}

H. A. Nguyen contributed to the methodology, formal analysis, writing of the manuscript, providing professional oversight and manuscript revision. A. Tura was responsible for conceptualization and manuscript revision.

\section*{Competing interests}
The authors declare no competing interests.

\section*{Data availability}
Access to the D-Wave System in this study is available at \href{https://cloud.dwavesys.com}{cloud.dwavesys.com}. The code related to this research be accessed at \href{https://github.com/x-repos/traveltime-inversion-annealing}{https://github.com/x-repos/traveltime-inversion-annealing}. 





\end{document}